\documentclass[12pt,a4paper]{article}

\usepackage{graphicx}
\usepackage{bm}% bold math

\begin{document}
\title{Development of data-analysis software \\ for total-reflection high-energy positron diffraction (TRHEPD)}
\author{Kazuyuki Tanaka${}^1$\thanks{corresponding author e-mail D19T1101M@edu.tottori-u.ac.jp} , 
Takeo Hoshi${}^1$, Izumi  Mochizuki${}^2$, \\
Takashi Hanada${}^3$, Ayahiko Ichimiya${}^2$, Toshio Hyodo${}^2$ \\ %list of all authors
{\it\normalsize ${}^1$Department of Applied Mathematics and Physics, Tottori University}\\ %first affiliation first line
{\it\normalsize   4-101 Koyama-Minami, Tottori 680-8550, Japan}  \\ %first affiliation second line
{\it\normalsize ${}^2$ Institute of Materials Structure Science, High Energy Accelerator Research Organization }\\ %secopnd affiliation first line
{\it\normalsize 1-1 Oho, Tsukuba, Ibaraki 305-0801, Japan} \\ %first affiliation second line etc.
{\it\normalsize ${}^3$Institute for Materials Research, Tohoku University, }\\ %secopnd affiliation first line
{\it\normalsize 2-1-1 Katahira, Aoba-ku, Sendai 980-8577, Japan} }%first affiliation second line etc.
\maketitle

\noindent
\begin{abstract}
%\underline{abstract}
The present paper reports on the recent activity of 
the data analysis software development  
for total-reflection high-energy positron diffraction (TRHEPD), 
a novel experimental technique for surface structure determination.
Experiments using TRHEPD are being conducted intensively at the Slow Positron Facility, 
Institute of Materials Structure Science, High Energy Accelerator Research Organization, 
revealing surface structure of interest. 
The data analysis software
provides a solution to the inverse problem in which the atomic positions of a surface structure are 
determined from the experimental diffraction data (rocking curve). 
The forward problem  is solved by the numerical solution of the partial differential equation in the quantum scattering problem.  
A technical demonstration with a test problem was carried out to confirm the software functioned as expected. 
Since the analysis method has a general mathematical foundation, 
it is also applicable to other experiments, such as X-ray or electron diffraction experiments.
\end{abstract} 
  
\section{Introduction}

Since material properties  are governed by 
the atomic structure or 
the type and position of each atom, 
the information of the structure is critical in discerning these properties. 
For bulk structure, X-ray diffraction is 
the standard technique used to determine the structure of the crystals of new materials, proteins, and so on. 
For surface structure, however, a standard technique for the definitive determination of the atomic structure 
of the topmost and subsurface atoms is not yet established. 
%This has resulted in a number of studies where possible atomic geometries consistent with the observed characteristics are proposed; 
%as a consequence, sometimes many different models are proposed by various authors.

Total-reflection high-energy positron diffraction (TRHEPD) has been developed as 
a novel method for such surface structure determination. 
At the Slow Positron Facility (SPF), 
Institute of Materials Structure Science (IMSS), High Energy Accelerator Research Organization (KEK) much work has been conducted, successfully
and revealing surface structures of the surfaces of interest. 
(see a review~\cite{FUKAYA2018-JPHYSD}).
For example, TRHEPD determined the structure of the rutile-TiO$_2$(110) (1 $\times$ 2) surface 
which had been under debate over 30 years \cite{Mochizuki2016_TiO2}. 

Here, we report recent activity on the software development for 
the analysis of TRHEPD data. 
It is based on the inverse problem in which  
the surface structure is determined  from the experimental diffraction data. 

The present paper is organized as follows:
experiment and theory of TRHEPD is explained briefly in Section ~\ref{SEC-TRHEPD}; 
overview of the data analysis software is 
presented in 
Section \ref{SEC-DATA-ANALYSIS-METHOD};
Section ~\ref{SEC-DEMO} details the 
technical demonstration of the software and 
associated discussions;
a summary and a future aspect is given in Section \ref{SEC-SUMMARY}.
%for the reproduction of a known surface structure. 
%Discussion 

%Here, 
%$X=(X_1,X_2, \cdot \cdot \cdot ,X_N)$ indicates
%the position of the surface and 
%the atomic positions
%It is based on the inverse problem in which  
%of a surface structure are determined 
%from the experimental diffraction data $D_{\rm exp} (D_{\rm exp} \Rightarrow X)$.

%It is based on the inverse problem in which the atomic positions $X$ is determined
%from the diffraction 

%%%%%%%%%%%%%%%%%%%%%%%%%%%%%%%%%%%%%%%%%%%%%%%%%%
\begin{figure}[h]
\begin{center}
  \includegraphics[width=0.4\textwidth]{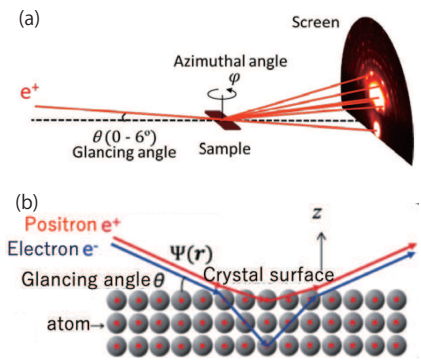}
\end{center}
\caption{Schematic diagram of (a) experimental setup and (b) typical paths of the beams are schematically depicted both 
for the positron  (THREPD) and electron (RHEED) cases.
}
\label{FIG-THREPD-OVERVIEW}       % Give a unique label
\end{figure}
%%%%%%%%%%%%%%%%%%%%%%%%%%%%%%%%%%%%%%%%%%%%%%%%%%

\section{TRHEPD \label{SEC-TRHEPD} }

Features of TRHEPD 
are shown schematically 
in Figure \ref{FIG-THREPD-OVERVIEW} and 
will be briefly explained in this section. 
The experimental basis is the same as 
in reflection high-energy electron diffraction (RHEED) experiments. 

The problem is to determine
the positions of the atoms at the top-most surface layer and several subsurface layers. 
Hereafter, 
the $z$ axis is assumed to be perpendicular to the material surface
and the coordinates are denoted as $\bm{r}=(x,y,z)$.
$N_{\rm a}$ is the number of the atoms of which positions 
$((x_i,y_i,z_i), i=1,2,...,N_{\rm a})$ will be determined through the data analysis. 

The data analysis provides a solution to the inverse problem in which  
the surface structure, $X$, is determined from the experimental diffraction data 
$D_{\rm exp} (D_{\rm exp} \Rightarrow X)$.
The calculated diffraction data $D_{\rm cal}$ 
is obtained from the position of the atoms $X$ \cite{ICHIMIYA1983}
and the calculation is called the forward problem $(X \Rightarrow D_{\rm cal}(X))$.

\subsection{Experiment}

%TRHEPD experiment 

The TRHEPD experiment is shown schematically in Figure \ref{FIG-THREPD-OVERVIEW}(a).
We also define the incident wave vector $\bm{K}^{\rm (in)} = (K \cos \theta \cos \varphi,K \cos \theta \sin \varphi, 0)$ projected on the $x$-$y$ plane,
where $K$ is the positron wave number in the vacuum.
The incident wave direction is characterized
by the glancing angle $\theta$ and the azimuthal angle $\varphi$. 
%We also define
%the  incident wave vector 
%$\bm{K}^{\rm (in)} = (K \cos \theta \cos \varphi,K \cos \theta \sin \varphi, 0)$
%projected on the $x$-$y$ plane.
 
Diffraction spots on the screen in Figure \ref{FIG-THREPD-OVERVIEW}(a)
are characterized by 
the two-dimensional indices $(p,q)$ of reciprocal lattice rods.
The intensity of the specific spots, $D_{pq}$, is observed 
as a function of the incident glancing angle, $\theta$ ($D_{pq}=D_{pq}(\theta)$), 
and called the rocking curve.
The rocking curve, $D_{pq}(\theta)$, depends also on the azimuthal angle ${\varphi}$,
which is fixed during a rocking curve measurement.
The spot with the indices of $(p,q)=(0,0)$ is usually brightest 
and called 00 spot. 
This paper focuses on the 00 spot and we will drop the indices for simplicity
($D=D(\theta)$). 
The observed data among discrete glancing angles is denoted
as $\bm{D} \equiv (D(\theta_1), D(\theta_2), ..., D(\theta_M))$,
where a typical number of the glancing angles is $M=$ 50 - 100. 
The present data analysis is carried out
by the normalized vector data $\bm{D}$ ($|\bm{D}|=1$)
and similarly normalized calculated values.

It is important that
the rocking curve with a given azimuthal angle ${\varphi}$
is hardly affected 
by the atomic coordinate parallel to the vector $\bm{K}^{\rm (in)}$.
If the vector $\bm{K}^{\rm (in)}$ is parallel to the $y$-axis, for example, 
the rocking curve depends only on the $x$ and $z$ components of the atomic position 
($\bm{D} = \bm{D}(x_1,x_2,...,x_{N_{\rm a}},z_1,z_2,...,z_{N_{\rm a}})$). 
In addition,
one can choose the azimuthal angle 
intentionally shifted from any high-symmetry zone axes
so that 
the rocking curve practically depends only on the $z$ components of the atomic position 
($\bm{D} = \bm{D}(z_1,z_2,...,z_{N_{\rm a}})$)
- this is called a one-beam condition \cite{ICHIMIYA1987}. 
Another choice for the azimuthal angle is that it is set along a high-symmetry zone axis
- this is called  a many-beam condition.  
These properties  allow us 
to reduce the number of variables in the data analysis from $3N_{\rm a}$ to $N_{\rm a}$.
Such dimensional reduction is of great advantage 
in realizing fast and reliable data analysis.  

Therefore, 
the measurement procedure, typically, consists of two stages,
where three diffraction data sets at different azimuthal angles are obtained. 
One data set is provided by an experiment in a one-beam condition, which is denoted as $\bm{D}^{\rm (OB)}$. 
The other two sets are provided by experiments in the many-beam condition, which are denoted as $\bm{D}^{\rm (MB1)}$ and $\bm{D}^{\rm (MB2)}$. 
%The incident wave vectors $\bm{K}^{\rm (in)}$ are orthogonal 
%between the two data sets in the many-beam  conditions. 
Mutually orthogonal incident wave vectors $\bm{K}^{\rm (in)}$ were chosen for the two data sets under the many-beam condition.
%The data analysis is carried out in a multi-stage procedure;
The first stage of the analysis procedure determines the $z$ component of the atomic position
$(z_1, z_2, ..., z_{N_{\rm a}})$ from the data set in the one-beam condition $(\bm{D}^{\rm (OB)})$.
The second stage determines the two components on the $x$-$y$ plane, $x$ and $y$ coordinates 
from the other two data sets each in the many-beam condition
$(\bm{D}^{\rm (MB1)},\bm{D}^{\rm (MB2)})$,
in which analysis the z components 
$(z_1, z_2, ..., z_{N_{\rm a}})$
determined in the first stage are fixed.

%As an advantage of TRHEPD,
%the $x$, $y$ or $z$ components of the atomic position 
%is determined independently from the experimental data with different azimuthal angle.
%In experiment, the azimuthal angle ${\varphi}$ should be chosen properly,
%since .

\subsection{Theory}

The theory or the forward problem $(X \Rightarrow D_{\rm cal}(X))$
of TRHEPD \cite{ICHIMIYA1983} is based on 
the quantum scattering problem 
of the positron wavefunction $\Psi(\bm{r})$;
the situation is shown schematically in Figure \ref{FIG-THREPD-OVERVIEW}(b). 
The partial differential equation with a given glancing angle and an azimuthal angle 
\begin{eqnarray}
\left( \Delta + K^2 + U(\bm{r})  \right) \Psi(\bm{r}) =   0
\label{EQ-PDE}
\end{eqnarray}
is solved numerically,
so as to obtain 
rocking curve data $\bm{D}=\bm{D}_{\rm cal}(X)$. 
%so as to reproduce the multiple-scattering effect \cite{ICHIMIYA1983}．
Here $U(\bm{r})$ is the crystal potential determined
by the atomic positions $X$.
The crystal potential $U(\bm{r})$ is periodic on the $x$-$y$ plane and can be
written by the two-dimensional Fourier series
\begin{eqnarray}
U(x,y,z) =   \sum_{m} U_{m}(z) \exp \left( i (k_x^{(m)} x+ k_y^{(m)} y) \right),
\label{EQ-POT}
\end{eqnarray}
where ($k_x^{(m)}, k_y^{(m)}$) 
is the surface reciprocal lattice vector of 
the $m$th rod ($p_m, q_m$).
%$U(\bm{r})$ is a complex function, owing to the absorption effect. 
In the numerical calculation \cite{ICHIMIYA1983}, 
the coordinate $z$ is discretized by the mesh grid with an equi-interval $h$ 
($z : = z_0 + j h, j=0,1,2,..$).
A typical value of the mesh interval $h$ is $h = 0.02$ \AA. 

The calculation under the one-beam condition 
\cite{ICHIMIYA1987} is much faster
than that under the many-beam condition,
since under the one-beam condition,
%the potential $U(x,y,z)$ is replaced by the one averaged in the $x$ and $y$ directions
%\begin{eqnarray}
%U(x,y,z) \Rightarrow U_{\rm ave}(z) \equiv \frac{1}{\Omega_{\rm cell}} %\int_{\rm cell} U(x,y,z) dx dy, 
%\label{EQ-POT2}
%\end{eqnarray}
%where $\Omega_{\rm cell}$ is the area of the two-dimensional unit cell.  
the only one Fourier component $(k_x^{(0)},k_y^{(0)})=(0,0)$ is non-zero in equation~(\ref{EQ-POT}) 
\begin{eqnarray}
U(x,y,z) = U_0(z). 
\label{EQ-POT-OB}
\end{eqnarray}
In this case, 
the wavefunction can also be written as $\Psi \equiv \Psi_0(z)$ and
equation ~(\ref{EQ-PDE}) is reduced to
a one-dimensional scattering problem 
\begin{eqnarray}
\left( \frac{d^2}{d z^2} + K^2 \sin^2 \theta + U_0(z)  \right) \Psi_0(z) =   0. 
\label{EQ-ODE}
\end{eqnarray}

It is noted that 
this calculation method \cite{ICHIMIYA1983} 
was originally developed 
for electron diffraction (RHEED)
but is applicable, 
since the TRHEPD method  
is different from that for RHEED
only in the sign of the incident-particle charge. 
As the electron beam penetrates into deeper layers than 
the positron, owing to the refraction off the surface, 
as schematically shown in Figure \ref{FIG-THREPD-OVERVIEW}(b), 
electron diffraction is less sensitive to surface structure than the positron diffraction.

%and the program code for the forward problem 

In the present paper, 
the forward problem is solved 
using the Fortran software developed by one of the authors 
 (A. Ichimiya) \cite{ICHIMIYA1983} and modified by another (T. Hanada) \cite{Hanada1995_RHEED}.
%so as to calculate $R=R(X)$ with a given (trial) atomic position $X$.  

\section{Data analysis method \label{SEC-DATA-ANALYSIS-METHOD} }

This section details 
the method used for the data analysis or the inverse problem  
$(\bm{D}_{\rm exp} \Rightarrow X)$. 
The inverse problem is solved by optimizing the residual function between
the calculated and experimental diffraction data
\begin{eqnarray}
R(X) \equiv | \bm{D}_{\rm cal}(X) - \bm{D}_{\rm exp}|
\label{EQ-R-FACTOR}
\end{eqnarray}
with a given experimental diffraction data $\bm{D}_{\rm exp}$, where $\bm{D}_{\rm cal}$ is also normalized 
($|\bm{D}_{\rm cal}|=1$).
The function $R(X)$ is called the reliability factor or R-factor. 

%%%%%%%%%%%%%%%%%%%%%%%%%%%%%%%%%%%%%%%%%%%%%%%%%%
\begin{figure}[h]
\begin{center}
  \includegraphics[width=0.4\textwidth]{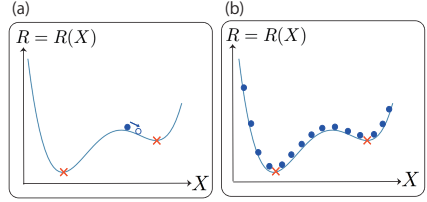}
\end{center}
\caption{Schematic figure for two optimization methods of 
the function $R=R(X)$: 
(a) local search or the iterative optimization;
(b) global search or 
simple-comparison optimization on grid.  
}
\label{FIG-ALGORITHM}       % Give a unique label
\end{figure}
%%%%%%%%%%%%%%%%%%%%%%%%%%%%%%%%%%%%%%%%%%%%%%%%%%

In general, 
optimization algorithms are classified
into local or global searches.
Two typical algorithms are illustrated in Figure ~\ref{FIG-ALGORITHM};
Figure ~\ref{FIG-ALGORITHM}(a) shows a local search algorithm
with iterative updates, in which
an initial structure data $X^{(0)}$ is prepared and
the data, $X$, is updated iteratively
($X^{(0)} \Rightarrow X^{(1)} \Rightarrow X^{(2)}  .. \Rightarrow X^{(i)} \Rightarrow...$), 
so as to decrease $R(X)$. 
Figure ~\ref{FIG-ALGORITHM}(b), on the other hand,
shows a global search algorithm
with a straightforward calculation on a mesh grid,
which will be discussed later in Section ~\ref{SEC-DISCUSSION}.
%The function $R=R(X)$ is calculated on the global mesh grid for the coordinate space of $X$. 

The present paper focuses
on a local search algorithm. 
The iterative optimization is realized using the
Nelder-Mead algorithm 
\cite{Nelder-Mead1965, Lagarias1998},
a commonly used gradient-free optimization algorithm,
where 
the gradient ($\nabla R$) is not calculated in any way. 
The iterative process is stopped
when it reaches a convergence criteria chosen by the user.
In this study, we developed a Python-based data analysis software.
%Python is the {\it de facto} standard programing language 
%in data driven science. 
The Nelder-Mead algorithm is realized
by the module in the scipy library (\verb|scipy.optimize.fmin|). 
The method is standard and the use of the scipy library is not essential. 

%%%%%%%%%%%%%%%%%%%%%%%%%%%%%%%%%%%%%%%%%%%%%%%%%%
\begin{figure}[h]
\begin{center}
  \includegraphics[width=0.3\textwidth]{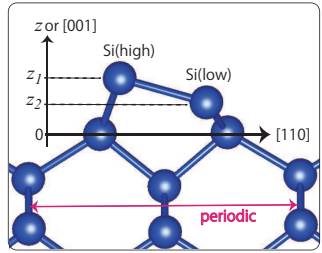}
\end{center}
\caption{Atomic positions of Si(001)-2 $\times$ 1 surface
with an asymmetric surface dimer.
The vertical axis ($z$ axis) is along the $[001]$ direction,
while the horizontal axis is along the $[110]$ direction. 
The periodic unit in the $[110]$ direction is indicated by the red arrowed line. 
}\label{FIG-SURFACE}       % Give a unique label
\end{figure}
%%%%%%%%%%%%%%%%%%%%%%%%%%%%%%%%%%%%%%%%%%%%%%%%%%

%%%%%%%%%%%%%%%%%%%%%%%%%%%%%%%%%%%%%%%%%%%%%%%%%%
\begin{figure}[h]
\begin{center}
  \includegraphics[width=0.3\textwidth]{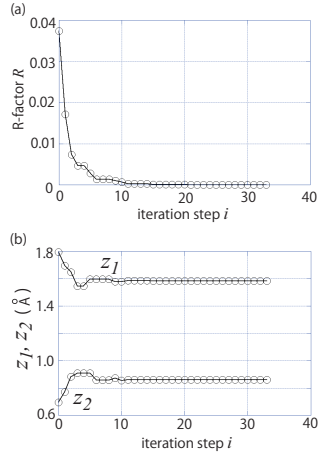}
\end{center}
\caption{
The data analysis for Si(001)-$2 \times 1$ surface: 
(a) The R-factor $(R=R(z_1,z_2))$ and 
(b) the $z$ coordinate of the surface atoms $(z_1, z_2)$
are plotted as the function of the iteration step $i$. 
}
\label{FIG-NM-CONVERGENCE}       % Give a unique label
\end{figure}
%%%%%%%%%%%%%%%%%%%%%%%%%%%%%%%%%%%%%%%%%%%%%%%%%%

%%%%%%%%%%%%%%%%%%%%%%%%%%%%%%%%%%%%%%%%%%%%%%%%%%
\begin{figure}[h]
\begin{center}
  \includegraphics[width=0.3\textwidth]{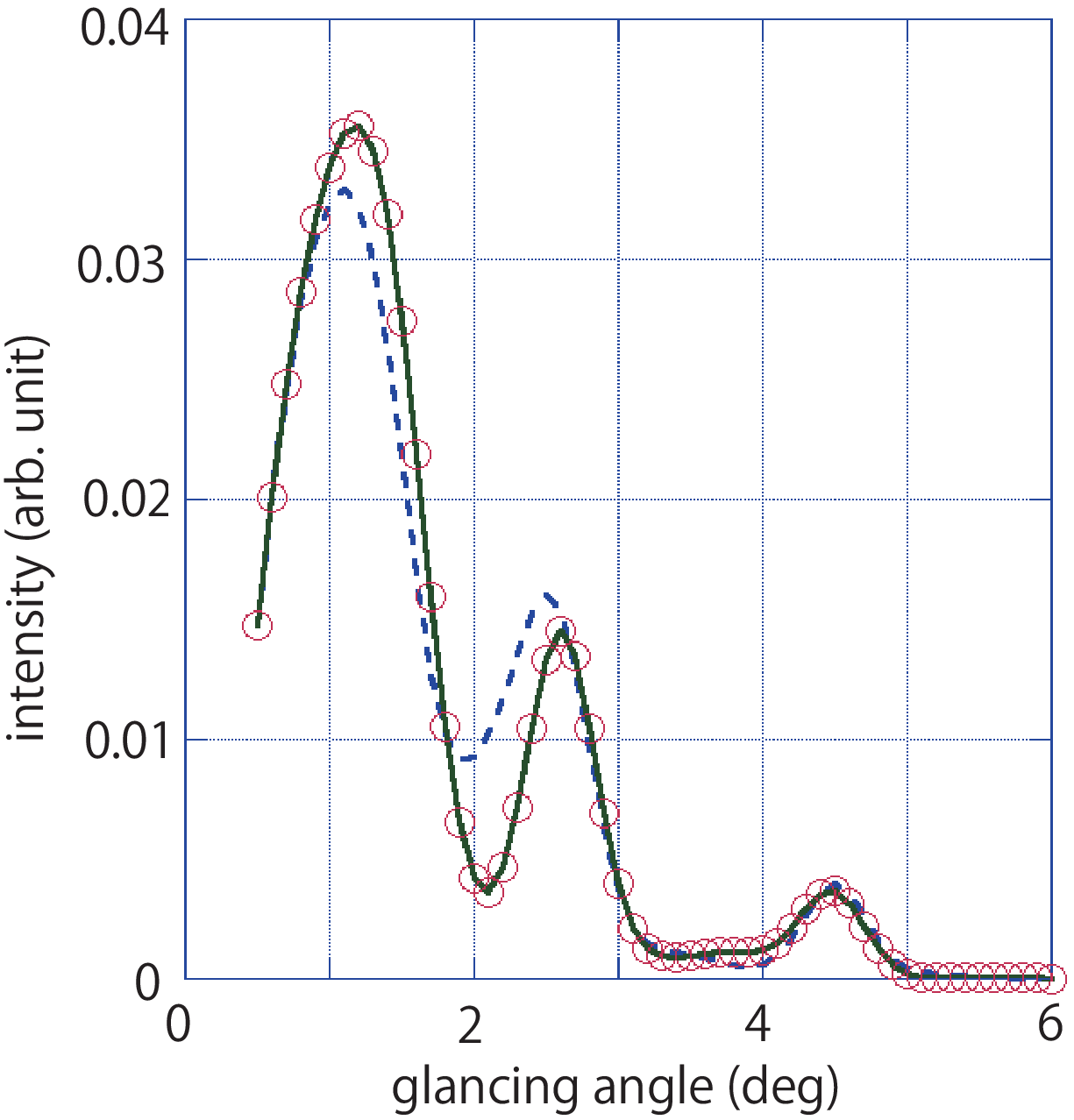}
\end{center}
\caption{
The rocking curves of the Si(001)-$2 \times 1$ surface
of the initial and final structures 
are plotted by dashed and solid lines, respectively.
The rocking curve in the reference structure 
is also plotted by circles. 
}
\label{FIG-ROCKING-CURVE}       % Give a unique label
\end{figure}
%%%%%%%%%%%%%%%%%%%%%%%%%%%%%%%%%%%%%%%%%%%%%%%%%%

\section{Technical demonstration  and discussion \label{SEC-DEMO}}

\subsection{Technical demonstration}
 
A numerical test problem was run
for the technical demonstration of our data analysis software.
The test problem used was to reproduce  
the Si(001)-$ 2 \times 1$ surface structure.
Studies of this structure, shown in Figure \ref{FIG-SURFACE},
have been reported in a number of papers, such as Reference \cite{STEKOL2002}.
The two top-most silicon atoms 
form an asymmetric dimer,
where the $z$ coordinate values of these atoms differ significantly.
The higher (vacuum-side) atom and the lower (bulk-side) atom
of the surface dimer are denoted as Si(high) and Si(low), respectively.
The asymmetry is induced to decrease total surface energy by the electron transfer from Si(low) to Si(high), 
which results in a lone pair 
in the dangling-bond state at Si(high) and 
an empty dangling-bond state at Si(low).
The $z$ coordinates of the higher and lower atoms
are denoted by $z_1$ and $z_2$ in Figure \ref{FIG-SURFACE}, respectively, 
where the origin of the $z$ axis ($z =0$) 
is located at the plane of the atoms in the second surface layer. 

In the present demonstration, 
the values of the $z$ component of the two top-most silicon atoms
$(z_1,z_2)$ are determined from the diffraction data in the one-beam condition.
A numerically generated \lq reference' data $\bm{D}_{\rm ref}$ is used, 
instead of a real experimental data $\bm{D}_{\rm exp}$.  
The reference data $\bm{D}_{\rm ref}$ is generated numerically 
for the known structure $(z_1, z_2) = (z_1^{\rm (ref)}, z_2^{\rm (ref)})$  
($\bm{D}_{\rm ref} \equiv \bm{D}_{\rm cal}(z_1^{\rm (ref)}, z_2^{\rm (ref)}$)) and
the R-factor is defined as
\begin{eqnarray}
R(z_1, z_2) \equiv | \bm{D}_{\rm cal}(z_1, z_2) - \bm{D}_{\rm cal}(z_1^{\rm (ref)}, z_2^{\rm (ref)})|,
\label{EQ-R-FACTOR-DEM}
\end{eqnarray}
where $(z_1^{\rm (ref)}, z_2^{\rm (ref)})$ = (1.5832 \AA, 0.8603 \AA).  
A demonstration with the numerically generated 
reference data was carried out 
in order to confirm that the Nelder-Mead algorithm was suitable and 
reached the exact solution $((z_1, z_2) \rightarrow (z_1^{\rm (ref)}, z_2^{\rm (ref)}))$. 
When analyzing with experimental data, 
an optimized value of $R \le 10^{-2}$ is usually acceptable 
in  surface structure determination research.

%%%%%%%%%%%%%%%%%%%%%%%%%%%%%%%%%%%%%%%%%%%%%%%%%%

The analysis results are summarized in Figures \ref{FIG-NM-CONVERGENCE} and \ref{FIG-ROCKING-CURVE}.
Figure  \ref{FIG-NM-CONVERGENCE} shows
the iterative optimization process.
The R-factor $(R=R(z_1,z_2))$ and 
the $z$ coordinates of the surface atoms $(z_1, z_2)$
are plotted as functions of the iteration step $i$,
in Figure \ref{FIG-NM-CONVERGENCE} (a) and (b), respectively.
Figures ~\ref{FIG-NM-CONVERGENCE} and \ref{FIG-ROCKING-CURVE} indicate that
the structure $(z_1, z_2)$ converges
to the exact solution $((z_1, z_2) \rightarrow (z_1^{\rm (ref)}, z_2^{\rm (ref)}))$ correctly. 
The R-factor changes from $R=0.037$  at the initial structure ($i=0$) 
into $R= 2 \times 10^{-6}$ at the final structure ($i=33$).
Figure \ref{FIG-ROCKING-CURVE} shows
the rocking curve $D_{\rm cal}=D_{\rm cal}(\theta)$ in the initial and final structure,
where the reference data $D_{\rm ref}=D_{\rm ref}(\theta)$ is also shown. 
Figure \ref{FIG-ROCKING-CURVE} confirms that
the atomic coordinates obtained by the search reproduce the reference data correctly. 
The analysis was carried out 
on a notebook PC with 
an Intel Core$^{\rm TM}$ i3-6006U processor
and the elapsed time was 72 s,
including  the computational time and the file I/O time.

It is of note that 
the difference in the atomic positions
$(z_1, z_2)$
between the initial and the final structures
is of the order of $0.1\,$ \AA $\,$ 
in Figure \ref{FIG-NM-CONVERGENCE}(b)  
and the difference can be clearly observed  
in the rocking curves of 
Figure \ref{FIG-ROCKING-CURVE}.
This demonstrates that  
TRHEPD 
can give an excellent selectivity 
or an ultra-fine spatial resolution 
in the order of $0.1\,$ \AA,
if the experimental uncertainty of the observed data is sufficiently small.

\subsection{Discussion \label{SEC-DISCUSSION}} 

A number of issues are discussed below on the current and prospective versions of the data analysis software.
(I) The present paper has detailed a  technical demonstration of the software. 
Data analysis with real experiment is ongoing. 
(II) The forward problem solver code used is planned to be parallelized for faster computation.
Since the computations of the intensity 
$\{ D_{\rm cal}(\theta_i) \}$ with
different $i$ numbers
are independent,
the procedure is ideal for parallelism on current  computers.
(III) Work is underway to develop further software with the global search algorithm shown in Figure \ref{FIG-ALGORITHM}, so as to avoid the local, not absolute, optimization. 
The software presently used provides an iterative optimization algorithm and requires an initial guess for the structure.
As such,
the analysis is based on a local search algorithm 
and can be trapped by a local minimum of $R(X)$. 
A simple-comparison algorithm may require significant computational cost because the number of sets of atomic coordinates increases exponentially with the increasing number of atoms and in reducing the mesh in the coordinates.  However, this type of algorithm is suitable for modern massive parallel supercomputers as the calculations of $R=R(X)$ among different atomic positions, $X$, of different atoms 
are independent procedures. 
Our global search software is now being tested on several supercomputers, such as the Oakforest-PACS supercomputer in Japan.   
(IV) In addition, we plan to develop another global search method
based on Besian inference with a Monte Carlo (stochastic) algorithm. 
The Monte Carlo algorithm is known as a reliable and efficient sampling method.
The method provides the posterior probability of atomic positions through Bayes' theorem and
enables us to evaluate the uncertainty of estimated atomic positions. 
This method has been applied to surface structure analysis by X-ray diffraction experimentation
\cite{ANADA2017}. 

\section{Summary and future aspect \label{SEC-SUMMARY} } 

A data analysis software is being developed for surface structure analysis 
by the total-reflection high-energy positron diffraction (TRHEPD).
The software provides a solution to the inverse problem,
where the forward problem is a quantum scattering problem or partial differential equation.
The software has a solid mathematical foundation and shows promise  
for the analysis of real experimental data.
The program code will be available online in the near future.

As a future aspect,
the software will be promoted to address general data analysis of surface structures,
not only for positron diffraction but also for X-ray  and electron diffraction
by simply changing the forward problem solver.

\section*{Acknowledgement}
The present research is supported by the Japanese government in the post-K project and KAKENHI funds (19H04125, 17H02828). Several numerical computations were carried out by the supercomputer Oakforest-PACS supercomputer for the Initiative on Promotion of Supercomputing for Young or Women Researchers, Information Technology Center, The University of Tokyo, the HPCI project (hp190066) and Interdisciplinary Computational Science Program in the Center for Computational Sciences, University of Tsukuba. Several computations were carried out also on the facilities of the Supercomputer Center, the Institute for Solid State Physics, the University of Tokyo and the Academic Center for Computing and Media Studies, Kyoto University.

\end{document}